\documentclass[a4paper]{jpconf}
\usepackage{graphicx}
\newcommand{\xbj}{x_B}
\newcommand{\ph}{\phi_h}
\newcommand{\zh}{z}

\usepackage[top=2.5cm, bottom=2.5cm, left=2.5cm, right=2.5cm]{geometry}

\begin{document}
\title{Studies of spin-orbit correlations at JLAB}

\author{Mher Aghasyan$^1$, Harut Avakian$^2$ for the CLAS collaboration}

\address{$^1$ INFN-LNF, Via E. Fermi 40, Frascati (RM), 00044, Italy}
\address{$^2$ Thomas Jefferson National Accelerator Facility, Newport News, Virginia 23606, USA}

\ead{mher@jlab.org}

\begin{abstract}
Studies of single spin asymmetries for pion electroproduction in semi-inclusive 
deep-inelastic scattering are presented  using 
 the polarized $\sim$6 GeV electrons from at the Thomas Jefferson National Accelerator
Facility (JLab) and the Continuous Electron Beam Accelerator Facility (CEBAF) Large Acceptance 
Spectrometer (CLAS) with the Inner Calorimeter. 
The cross section versus the azimuthal angle $\phi_h$ of the produced neutral
pion has a substantial $\sin\phi_h$ amplitude. The dependence of 
this amplitude on Bjorken $x_B$ and on the pion transverse
momentum is extracted and compared with published data.  
\end{abstract}

\section{Introduction}
In recent years  it became clear that understanding the orbital motion of partons, and in 
particular the role of partonic initial and final state interactions of quarks, is crucial 
for the construction of a more complete picture of the nucleon in terms of elementary quarks 
and gluons. 
Parton distribution functions have been generalized to contain information not only on the 
longitudinal momentum but also on  the transverse momentum distributions of partons in a fast moving hadron.
Intense investigation of Transverse Momentum Dependent (TMD) distributions of partons both from the
experimental and theoretical sides, indicate that QCD-dynamics inside hadrons is  
much richer than what can be learned from collinear parton distributions.

Two fundamental mechanisms have been identified leading to  single-spin asymmetries 
(SSAs) in hard processes:
the Sivers mechanism \cite{Sivers:1990fh,Anselmino:1998yz,Brodsky:2002cx,Collins:2002kn,Ji:2002aa}, which generates 
an azimuthal asymmetry in the distribution of quarks in the nucleon due to their orbital motion,
and the Collins
mechanism \cite{Collins:2002kn,Mulders:1995dh}, which generates 
an asymmetry  during the hadronization of quarks.
TMD distributions  also contain unique information on the role of initial and final state
interactions of active partons in hard scattering processes~\cite{Brodsky:2002cx,Collins:2002kn,Ji:2002aa}.

Semi-inclusive deep-inelastic scattering (SIDIS) has emerged as a powerful tool to probe 
nucleon structure and provide direct access to TMDs through measurements of spin and azimuthal asymmetries.
In particular, SSAs,  arising from correlations of the  transverse parton momentum and  
the transverse spin of the parton or from the initial or final state hadron, provide unprecedented information 
about spin-orbit correlations.
QCD factorization for 
semi-inclusive deep-inelastic scattering at low transverse momentum in the 
current-fragmentation region has been established in references~\cite{Ji:2004wu,
Collins:2004nx, Bacchetta:2008xw}.  This new framework provides a rigorous basis to study the 
TMD parton distributions from SIDIS data using different spin-dependent and 
independent observables.  The analyses of the TMDs strongly depends also on knowledge of the fragmentation functions ~\cite{deFlorian:2009vb,Amrath:2005gv,Bacchetta:2007wc,Matevosyan:2010hh}.
Many experiments are currently trying to pin down various 
TMD effects through 1) semi-inclusive deep-inelastic scattering (  
HERMES at DESY~\cite{Airapetian:1999tv,Airapetian:2001eg,Airapetian:2004tw,Airapetian:2006rx}, 
COMPASS 
at CERN~\cite{Alexakhin:2005iw}, CLAS and Hall-A at Jefferson 
Lab~\cite{Avakian:2003pk,Avakian:2005ps,Avakian:2010ae}), 
2) polarized proton-proton collisions (PHENIX, STAR and BRAHMS at RHIC
 \cite{Adams:2003fx,Chiu:2007zy,Arsene:2008mi}) 
and 
3) electron-positron annihilation (Belle at KEK~\cite{Abe:2005zx}).

The cross section for single pion production
by longitudinally polarized electrons 
scatteried from unpolarized protons may be written in terms of a
set of structure functions
\cite{Mulders:1995dh,Levelt:1994np,Bacchetta:2006tn}. 
The helicity 
 dependent part ($\sigma_{LU}$) 
arises from the anti-symmetric part of the hadronic tensor:

\begin{eqnarray}
\label{HLT}
\frac{d\sigma_{LU}}{d\xbj dy,d\zh d^2P_T d\ph}  \propto \lambda_e \times \sqrt{y^2+\gamma^2} \sqrt{1-y-\frac{1}{4}\gamma^2}  \sin \ph\,\,{F^{\sin \ph}_{LU}}.
\end{eqnarray}

\noindent The subscripts on $\sigma_{LU}$ specify the beam and  target 
polarizations, respectively ($L$ stands for longitudinally 
polarized and $U$ for unpolarized).
The  azimuthal angle $\ph$ is the angle between leptonic and hadronic 
plane according to the Trento convention \cite{Bacchetta:2004jz}.
The kinematic variables $\xbj$, $y$, and $z$  are defined as: 
$
\xbj = Q^2/{2(P_1q)}, \,\, y={(P_1q)/(P_1k_1)}, \,\, \zh={(P_1P)/(P_1q)}, 
$
where $Q^2=-q^2=-(k_1-k_2)^2$ is the four-momentum 
of the virtual photon, $k_1$ ($k_2$) is the four-momentum of the incoming (scattered) lepton,
$P_1$ and $P$ are the four momenta of the target nucleon and the observed final-state 
hadron (respectively), $\gamma^2=4M^2\xbj^2y^2/Q^2$, $M$ is the nucleon mass, and $\lambda_e$ is the electron beam helicity.

The structure function $F^{\sin \ph}_{LU}$ arises due to the 
interference of the longitudinal and transverse photon contributions.
In the partonic description
of SIDIS, 
 assuming factorization holds, contributions to structure functions can be written 
as convolutions of parton distribution and fragmentation
functions  dependent on the scaling variables $\xbj$ and $z$,
respectively~\cite{Mulders:1995dh,Ji:2004wu}. 

The beam-spin asymmetries
in single-pion production off an unpolarized target are higher-twist by nature 
\cite{Afanasev:2003ze,Yuan:2003gu}.
Higher-twist observables are important  for understanding the
long-range quark-gluon dynamics and
may also  be accessible as leading contributions through the measurements of
certain asymmetries 
\cite{Levelt:1994np,Jaffe:1991ra,Tangerman:1994eh,Kotzinian:1994dv},
in particular the beam SSAs. 

Recently, higher-twist effects in SIDIS were interpreted in terms of 
average transverse forces acting on the active quark at  the instant after
being struck by the virtual photon \cite{Burkardt:2008vd}.
Different contributions to 
the beam SSAs discussed so far provide information on leading and 
sub-leading parton
distribution and fragmentation functions, related both to Collins and 
Sivers production
mechanisms.
Sizable beam SSAs were predicted \cite{Yuan:2003gu} based on spin-orbit correlations 
as the dynamical origin.
Within this framework, the asymmetry generated at the
distribution level, 
is given by either the convolution of the T-odd parton distribution
$h_1^{\perp}$ with the twist-3 fragmentation function $E$ 
\cite{Jaffe:1991ra}, or the convolution of the  twist-3 T-odd distribution
function $g^\perp$ with the unpolarized fragmentation function $D_1$\cite{Metz:2004je}.

\section{The $\pi^0$ beam spin asymmetry } 

In this section we present measurements from the E01-113 CLAS dataset of 
beam-spin asymmetries in the electroproduction of
neutral pions in deep-inelastic scattering
using the 5.776 GeV
electron beam and the CEBAF Large Acceptance 
Spectrometer (CLAS) \cite{Mecking:2003zu} at Jefferson Laboratory.
Longitudinally polarized electrons are scattered off
a liquid-hydrogen target. 
The beam polarization, frequently measured  with a 
M{\o}ller polarimeter, was on average $0.80$ 
with a fractional systematic uncertainty of 3\%.
The beam helicity was flipped every 30 ms to minimize systematic instrumental effects.
Scattered electrons were detected in CLAS.
Electron candidates were selected by a hardware trigger using a 
coincidence between
the gas Cherenkov
counters and the lead-scintillator electromagnetic calorimeters (EC). 

\begin{figure}[h]
\begin{minipage}{18pc}
\includegraphics[width=18pc]{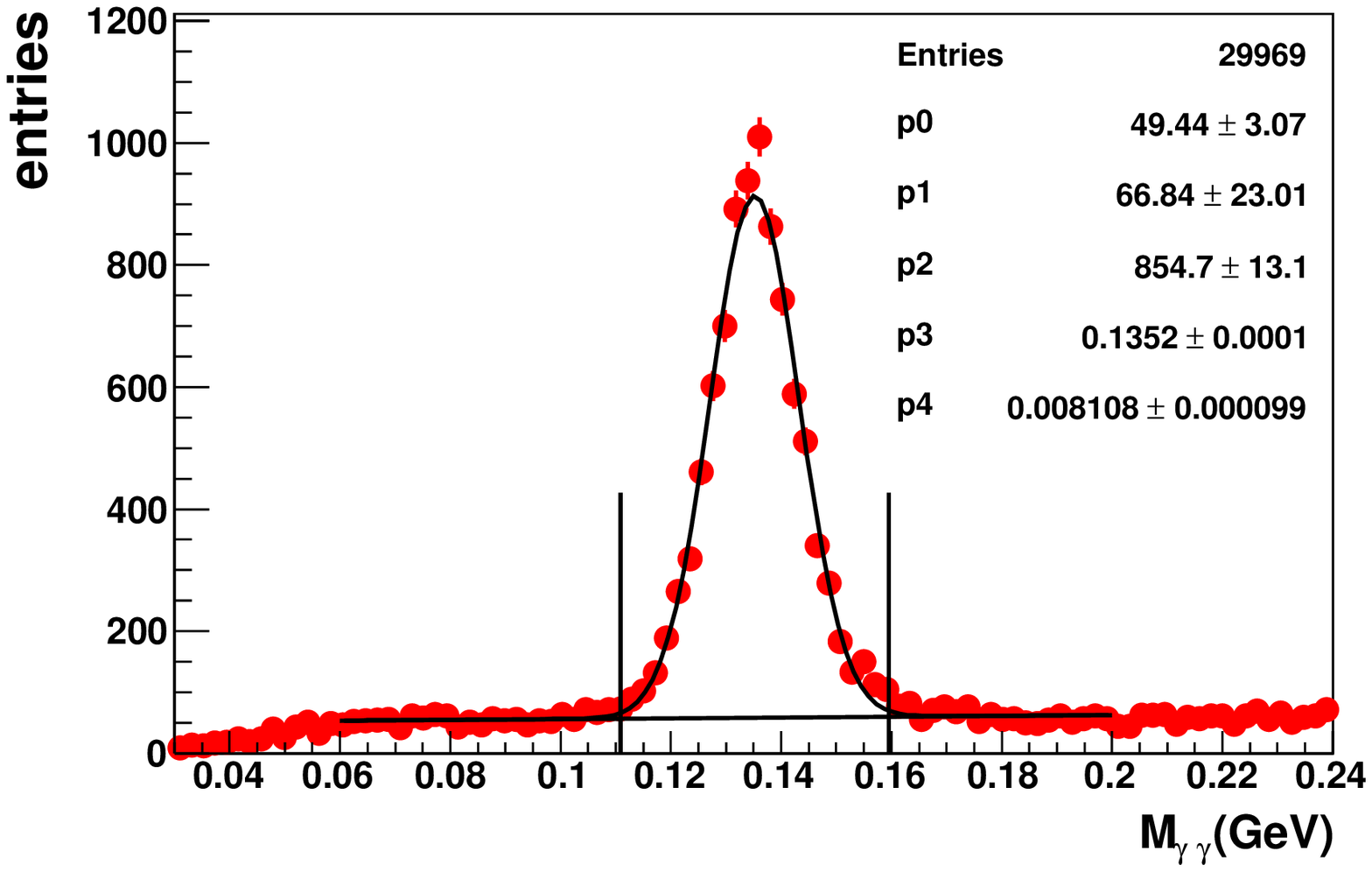}
\caption{\label{pi0bgkfit}Invariant mass spectrum of the $\gamma \gamma$ system $M(\gamma \gamma)$ in an arbitrarly chosen $\xbj$, $P_T$, $z$ and $\phi _{h}$-bin, fitted by gausian plus linear polinomial. Vertical black lines indicates $\pm 3\sigma$ from the mean.}
\end{minipage}\hspace{2pc}%
\begin{minipage}{18pc}
\includegraphics[width=18pc]{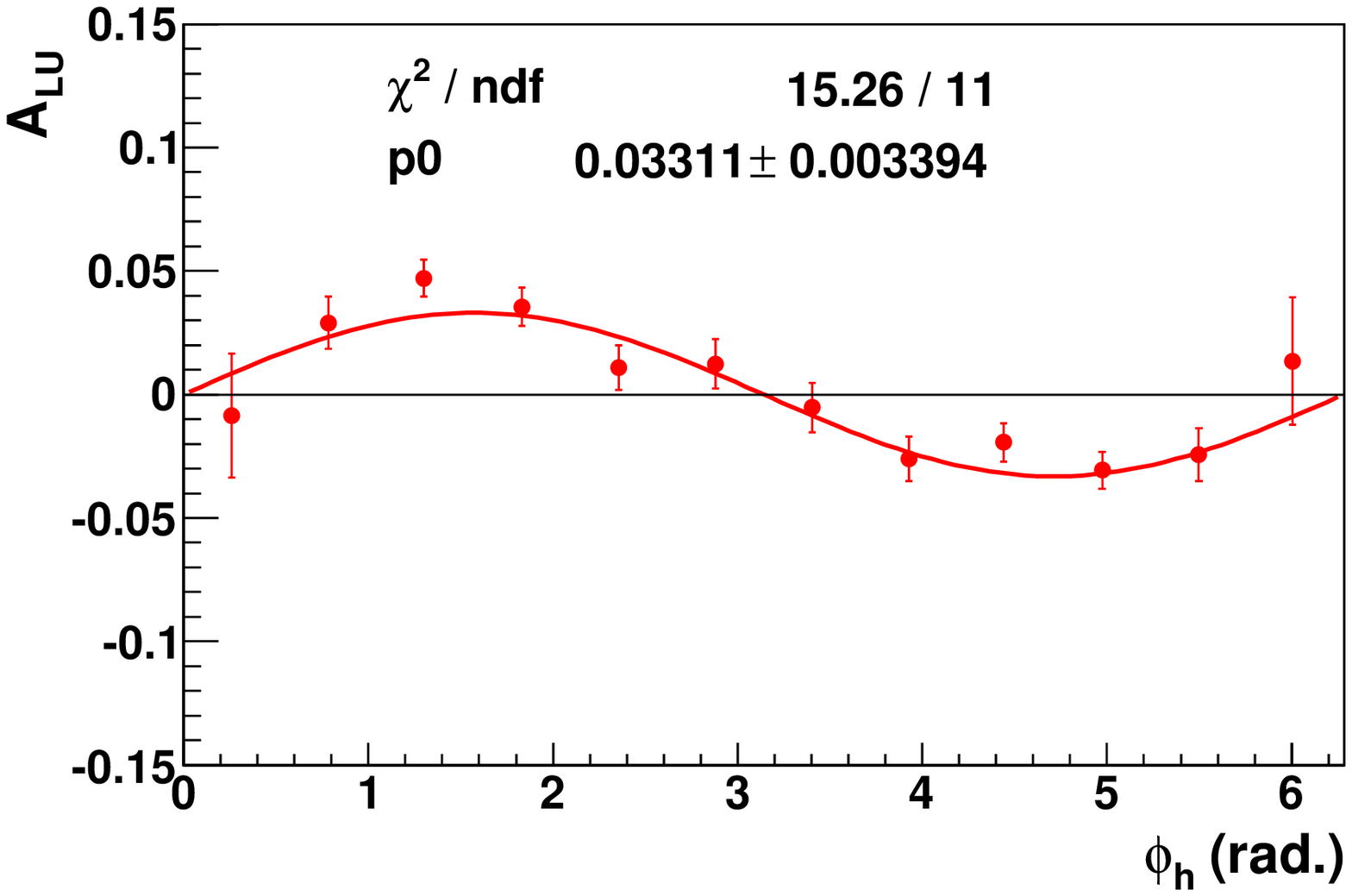}
\caption{\label{exFITsin12bin}Example of a $p_0 \cdot \sin(\ph)$ fit for $0.4<z<0.7$, $0.1<\xbj<0.2$ and $0.2$ GeV/c$ <P_T<0.4$ GeV/c. Only statistical error bars are shown.}
\end{minipage} 
\end{figure}

Neutral pion events were identified by calculating the invariant mass of two photons
detected with the CLAS electromagnetic calorimeter and the Inner Calorimeter (IC) \cite{girod:2007jq}.
In each kinematic bin, $\pi^0$ events are selected by a gaussian plus linear fit (see figure \ref{pi0bgkfit}). 
The combinatorial background is subtracted in each bin from the number of events inside $3 \sigma$ using the 
linear component of the fit.

Deep-inelastic scattering events are selected by requiring $Q^2>1$ GeV$^2$ and  $W^2>4$ GeV$^2$,
where $W$ is the invariant mass of the hadronic final state.
Events with low missing mass of the $e\pi^0$ system ($M_X<1.5$ GeV) were discarded to exclude
contributions from exclusive processes. 
The minimum value of the transverse $\pi^0$ momentum, $P_T>0.05$ GeV, ensures that the azimuthal
angle $\ph$ is well defined.
The total number of selected electron-$\pi^0$ coincidences  
is $\approx 3.0\times 10^6$.

The beam-spin asymmetry has been calculated for each kinematic bin as:
\begin{equation}
A_{LU} (\phi_{h})=\frac{1}{P} \frac{N^{+}_{\pi^0}(\phi_{h}) - N^{-}_{\pi^0}(\phi_{h})}{N^{+}_{\pi^0}(\phi_{h}) + N^{-}_{\pi^0}(\phi_{h}) }
\label{alueq0}
\end{equation}
where $P=0.794 \pm 0.024$ is the absolute beam polarization this data set and $N^{+}_{\pi^0}$ and $ N^{-}_{\pi^0}$ are the number of $\pi^0$'s with positive and negative beam helicity, respectively.
Asymmetry moments are extracted by
fitting the $\ph$-distribution of $A_{LU}$ in each $\xbj$ and $P_T$ bin with the theoretically motivated 
function $p_0 \cdot \sin(\ph)$.
An example of this fit is shown in figure~\ref{exFITsin12bin} for an arbitrarly chosen kinematic bin. 

\begin{figure*}[ht]
\begin{center}
\includegraphics[width=0.99\textwidth]{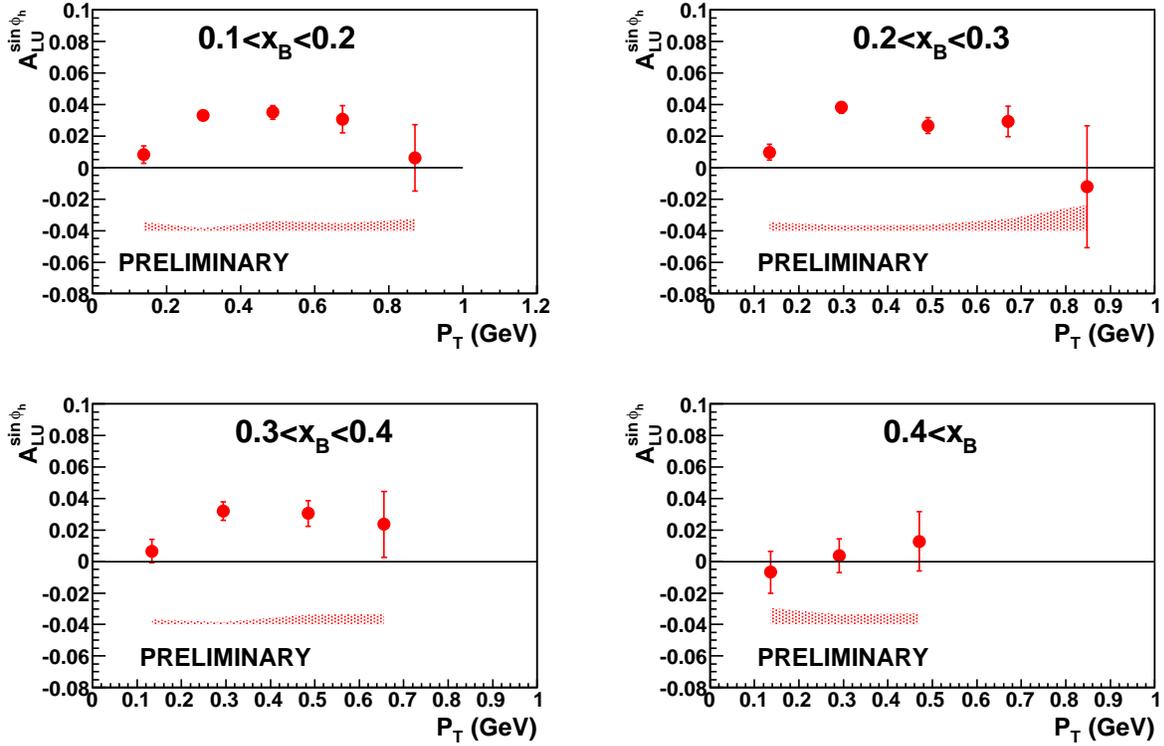}
\end{center}
\caption{$A^{\sin \phi_h}_{LU}$ as function of $P_T$ for different $\xbj$ ranges
and integrated over $0.4<z<0.7$. The error bars correspond to statistical and the bands to systematic uncertainties.
An additional 3\% uncertainty arises from the beam polarization measurement and another 3\% uncertainty 
from radiative effects.}
 \label{ALU_vs_PTXB_sysH}
\end{figure*}
In figure~\ref{ALU_vs_PTXB_sysH}, the
extracted $A_{LU}^{sin\phi}$ moment  is presented as a function of $P_T$ for different $\xbj$ ranges. 
Systematic uncertainties, presented by the bands, include the uncertainties due to the background 
subtraction, the event selection and possible contributions of higher harmonics in the extraction of the moments. These contributions are added in quadrature.
An additional 3\% uncertainty due to the beam polarization measurement and another 3\% uncertainty from radiative effects should be added to the systematic uncertainties.

The $A^{\sin \phi_h}_{LU}$ moment increases at low $P_T$ and reaches a plateau at values of about 0.3 GeV.
There is an indication that the decrease of $A^{\sin \phi_h}_{LU}$ at large $P_T$, 
expected from perturbative QCD, starts already at $P_T \sim$ 0.6 GeV. 

The surprising characteristics of favored and unfavored Collins functions, being of 
roughly equal magnitude but having opposite signs, as
indicated by latest measurements at HERMES \cite{Airapetian:1999tv,Airapetian:2004tw},
COMPASS \cite{Alexakhin:2005iw} and Belle \cite{Abe:2005zx},
puts the $\pi^0$ in a unique position in SSA studies.
Since the $\pi^0$ fragmentation function (FF) is the sum of $\pi^+$ and $\pi^-$ FFs, its
favored and unfavored contributions will be roughly equal and, in the case of the Collins FF, will cancel
each other to a large extent.
Contributions to the beam-SSA related to spin-orbit correlations can thus be studied without a significant 
background  from the Collins mechanism.

\begin{figure}
\begin{center}
\includegraphics[width=0.5\textwidth]{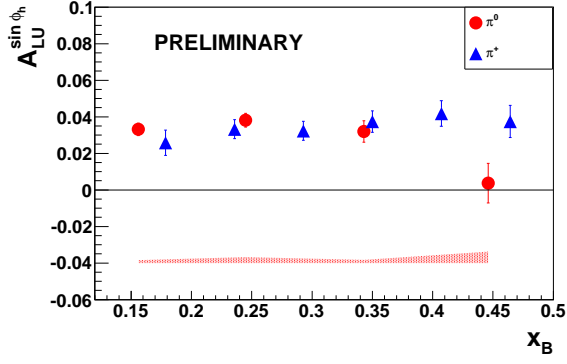}
\end{center}
\caption{The $\pi^0$ beam-spin asymmetry amplitude $A^{\sin \phi_h}_{LU}$ 
as function of $\xbj$ compared to that for $\pi^+$ from an earlier CLAS measurement~\cite{HAdubna}.
Uncertainties of the  $\pi^0$ measurement are as in figure~\ref{ALU_vs_PTXB_sysH}.
For both data sets $<P_T>\approx 0.38$ and $0.4<z<0.7$.}
\label{CLASaul}
\end{figure}

The measured beam-spin asymmetry amplitude for $\pi^0$ appears to be comparable  with the $\pi ^+$ asymmetry 
from a former CLAS data set \cite{HAdubna} both in magnitude and sign, as shown in figure~\ref{CLASaul}, 
indicating that contributions from the Collins mechanism cannot be the dominant ones.

\begin{figure}[h]
\begin{minipage}{18pc}
\includegraphics[width=18pc]{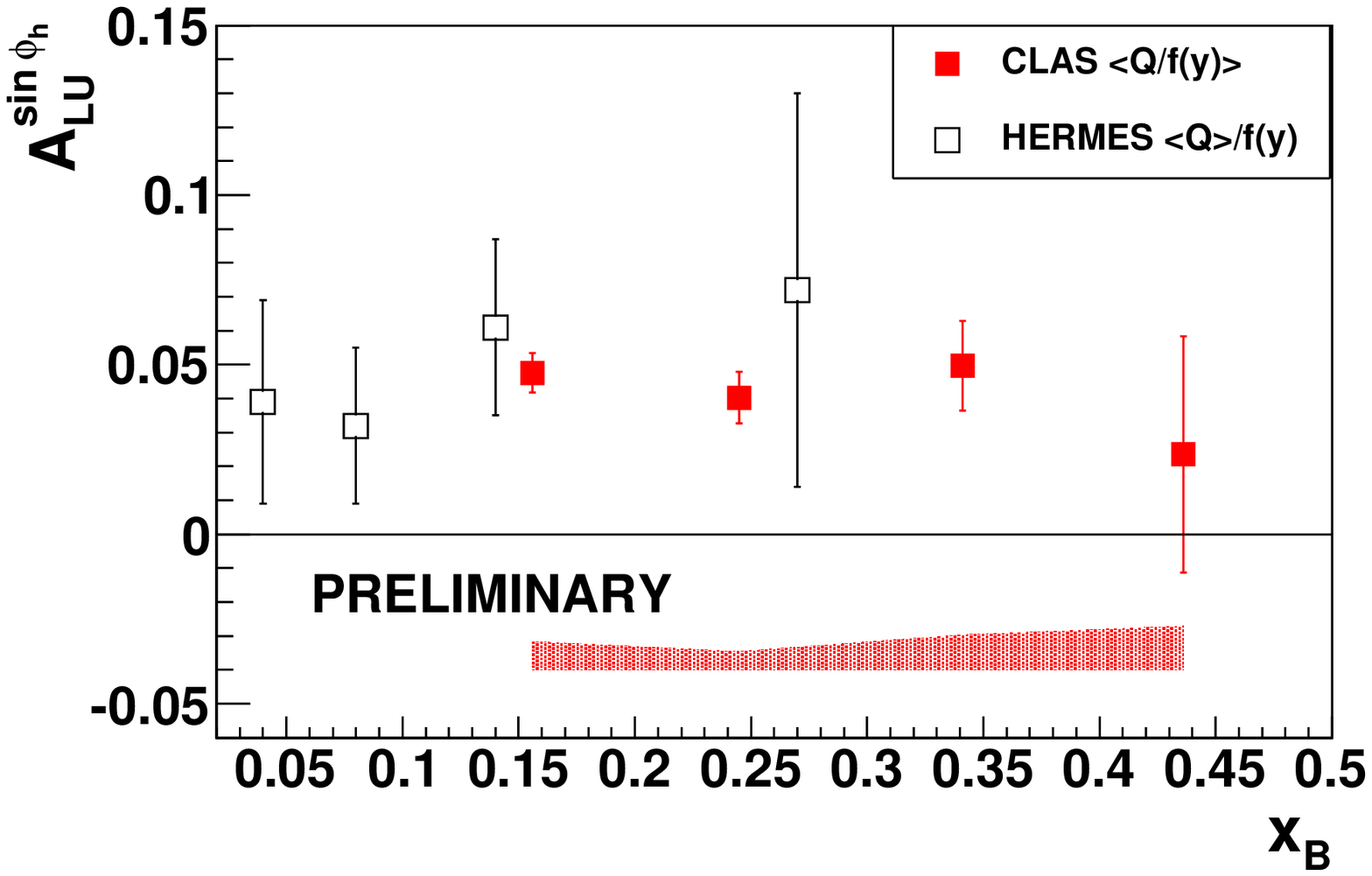}
\caption{\label{xbCH}$A^{\sin \phi_h}_{LU}$ multiplied  by the kinematic factor $<Q>/f(y)$ as a function of $\xbj$ from CLAS and HERMES \cite{Airapetian:2006rx}. The $0.4$ GeV/c$ <P_T<0.6$ GeV/c range for the CLAS data was used to compare with HERMES, since this is the closest kinematic range.}
\end{minipage}\hspace{2pc}%
\begin{minipage}{18pc}
\includegraphics[width=18pc]{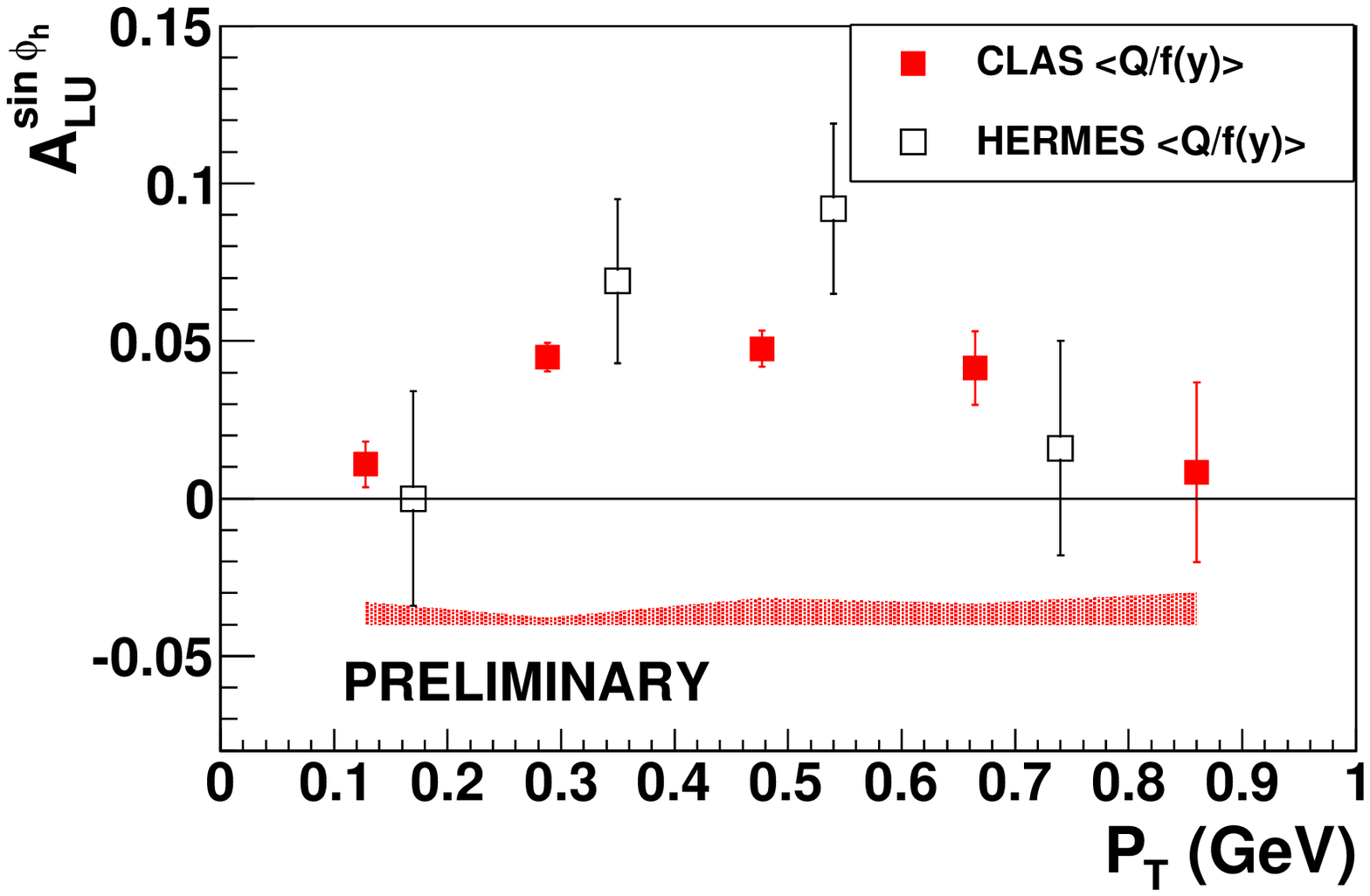}
\caption{\label{ptCH} $A^{\sin \phi_h}_{LU}$ multiplied by the kinematic factor $<Q>/f(y)$ as function of $P_T$  from CLAS and HERMES \cite{Airapetian:2006rx} (the same as in figure~\ref{xbCH}). The $0.1<\xbj<0.2$ range for the CLAS data was used to compare with, since this is the closest kinematic range.}
\end{minipage} 
\end{figure}

A similar measurement has been performed by the HERMES collaboration at a higher beam 
energy of 27.6 GeV~\cite{Airapetian:2006rx}. 
After taking into account the kinematic factors in the expression
of the beam-helicity dependent and unpolarized terms (\cite{Bacchetta:2006tn})
\begin{equation}
f(y)=\frac{y \sqrt{1-y}}{1-y+y^{2}/2} ,
\label{fy}
\end{equation}
 CLAS and HERMES measurements are found to 
be consistent as shown in figures~\ref{xbCH} and~\ref{ptCH}, indicating that at energies as low as 4-6 GeV the
behavior of beam-spin asymmetries is similar to higher energy measurements. The CLAS data provide
significant improvements in precision of beam SSA measurements in the kinematic region where the two data
sets overlap, and extend the measurements to the large $\xbj$ region not accessible by HERMES.

\subsection{Contamination from two-hadron production}

\begin{figure}[h]
\begin{minipage}{18pc}
\includegraphics[width=18pc]{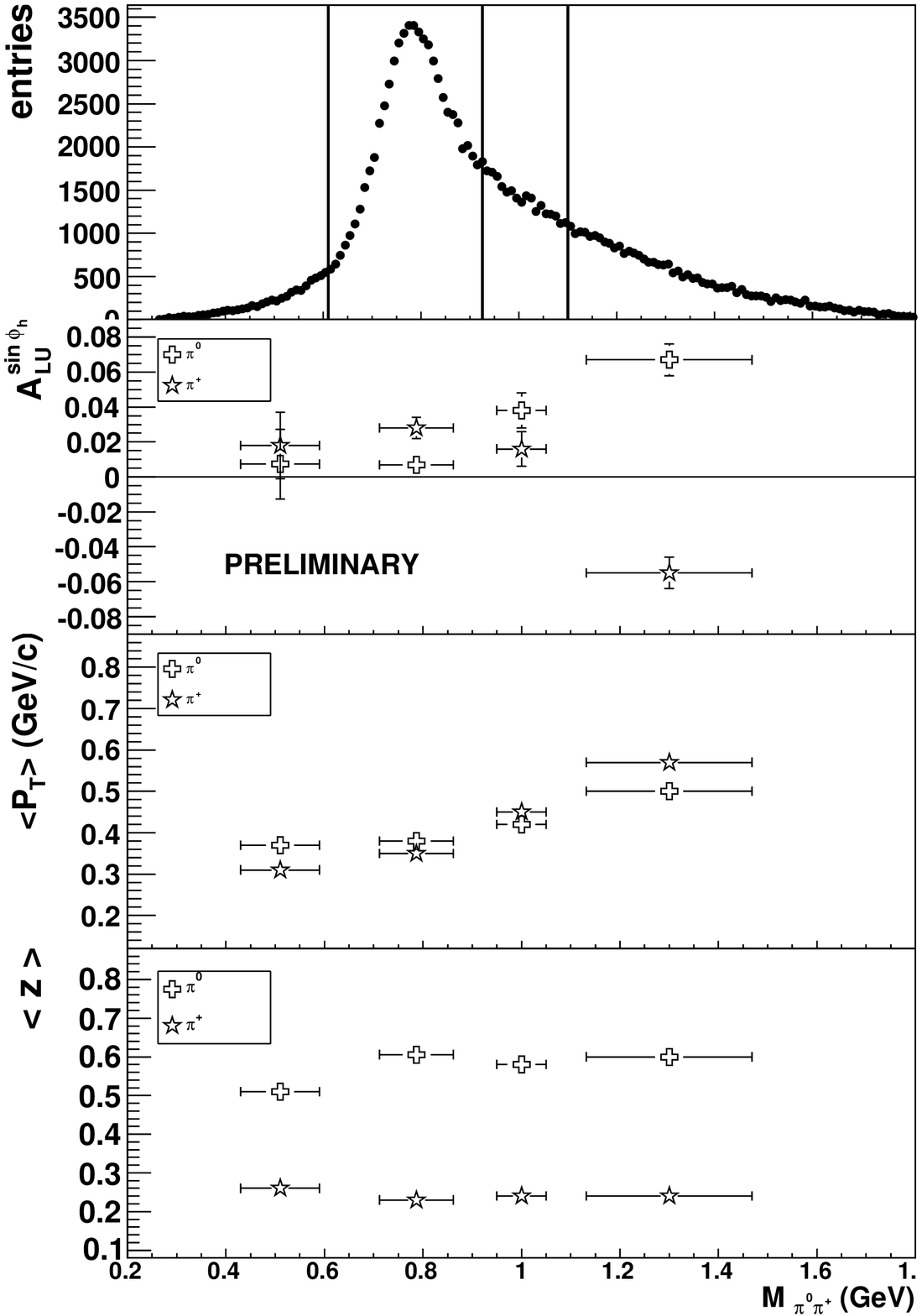}
\caption{\label{exrhop} From top to bottom: invariant mass $M_{\pi^0\pi^+}$ of two hadrons, $A^{\sin \phi_h}_{LU}$ , and average $<P_T>$ and $<\zh>$ as function of $M_{\pi^0\pi^+}$ for $\pi^0$ and $\pi^+$ in exclusive two-pion production.}
\end{minipage}\hspace{2pc}%
\begin{minipage}{18pc}
\includegraphics[width=18pc]{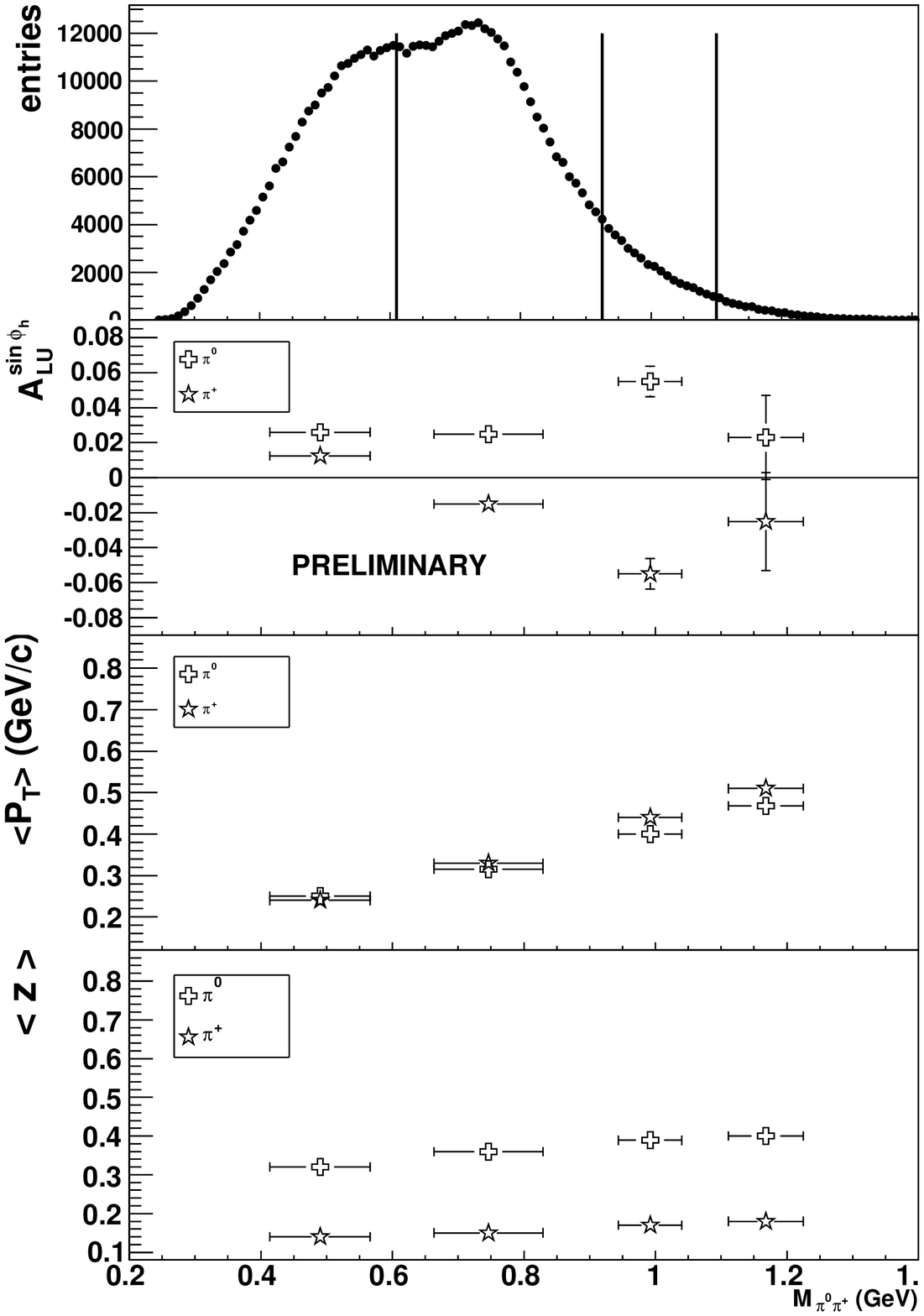}
\caption{\label{sirhop} From top to bottom: Invariant mass $M_{\pi^0\pi^+}$ of two hadrons, $A^{\sin \phi_h}_{LU}$ , and average $<P_T>$ and $\zh$ as function of $M_{\pi^0\pi^+}$ for $\pi^0$ and $\pi^+$ in non-exclusive production ($M_x(ep\rightarrow e'\pi^0 \pi^+ x)>1.5$ GeV).}
\end{minipage} 
\end{figure}

In this section a
 study of the beam spin asymmetry of single-pion production from exclusive and non-exclusive vector meson production is presented.
 In order to investigate a possible contamination of the presented results
by pions originating from the decay of exclusively produced vector mesons,
as advocated in~\cite{Airapetian:2006rx}, events with at least one $\pi^0$ and $\pi^+$ were selected from the same dataset. 
Exclusivity was ensured by requiring the missing mass $M_x$ of the 
$ep\rightarrow e'\pi^0 \pi^+ X$ system to be within 
0.8 GeV $< M_X(ep\rightarrow e'\pi^0 \pi^+ X)<1.1$ GeV. 
The invariant mass distribution of this exclusive two pion sample
is presented in the upper panel of figure~\ref{exrhop} where the exclusive
$\rho^+$ peak is clearly visible along with non-resonant exclusive
two-pion production.
The vertical lines indicate invariant mass ranges for which the 
beam spin asymmetry amplitude $A_{LU}^{\sin\phi_h}$ was extracted for each of the
two exclusively produced pions following the extraction method 
as described before.
These amplitudes are presented for $\pi^0$ and $\pi^+$ in the second top
panel of figure~\ref{exrhop} as labeled therein,  
along with the average values of $<P_T>$ and $<\zh>$ for the invariant mass
ranges.

Figure~\ref{sirhop} is the same as figure~\ref{exrhop} except for two-hadron events with $M_x(ep\rightarrow e'\pi^0 \pi^+ x)>1.5$ GeV (non exclusive).

The $A^{\sin \phi_h}_{LU}$ values for $\pi^0$($\pi^+$) increase (decrease) with increas invariant mass, and reach their maximum absolute value at the highest invariant mass. 

In contrast to a similar study for  $\rho^0$ presented in  \cite{Airapetian:2006rx}, the asymmetres for $\pi^0$ and $\pi^+$ have minimal values in the invariant mass range where $\rho^+$ events dominant, and furthermore the highest bin of invariant mass, the asymmetries for $\pi^0$ and $\pi^+$ have different signs. 
It should be noted also, that in these studies the momentum distributions and the acceptance for the two pions are not similar, while in \cite{Airapetian:2006rx} the two pions have similar momentum distributions and acceptance. $A^{\sin \phi_h}_{LU}$ of the SIDIS $\pi^0$ and $\pi^+$ in the same kinematic bin have the same sign and their amplitudes are roughly equal (figure~\ref{CLASaul}). For the case of two-hadron production (figure~\ref{sirhop}), in all invariant mass bins (except the first) $A^{\sin \phi_h}_{LU}$ for $\pi^0$ and  $\pi^+$ have  different signs, which is another reason for careful studies of two hadron production.

Comparison of the measured beam SSAs for pions originating from electroproduction of two-pions in
exclusive  and non-exclusive processes, suggests that the SSAs depend mainly on the kinematics of
the pion, in particular on $\zh$ and $P_T$, and are comparable for pions from different samples within
the same kinematical bin.

\section{Conclusions}

We have presented  measurements of kinematic dependences
 of the beam-spin asymmetry in semi-inclusive $\pi ^0$ electroproduction from the E01-113 CLAS dataset. 
The $\sin \phi_h$ amplitude is extracted as a function of $\xbj$ and the transverse pion 
momentum $P_T$, integrating over the $z$-range $0.4<z<0.7$.

The asymmetry shows no significant $\xbj$-dependence over the measured range.
The $P_T$ dependence of $A^{\sin \phi_h}_{LU}$ is consistent with an increase at low values of  $P_T$
which reaches a plateau for values $P_T >$ 0.3 GeV.
There is an indication that the decrease of $A^{\sin \phi_h}_{LU}$ at large $P_T$, expected from perturbative QCD, 
already starts at $P_T \sim$ 0.6 GeV. 
The observed asymmetry amplitudes for $\pi ^0$ indicate that the major contribution pion beam SSAs 
may be due to spin-orbit correlations. 

The results obtained are compared with published HERMES data~\cite{Airapetian:2006rx}, 
providing significant improvement in precision and an important 
input for studies of higher twist effects.
Despite the fact that the partonic formalism is much better suited
for higher energy reactions, there is a reasonable agreement in size and behavior
of  beam SSAs measured over a wide energy range.

Preliminary results on two-hadron production show interesting trends: opposite signs of asymmetries for $\pi^0$ and $\pi^+$, the highest asymmetry for the highest invariant mass range.   The beam spin asymmetry from $\pi^0$ and $\pi^+$ two-hadron production has been studied  for the first time as a function of the invariant mass of two pions. 
The $A^{\sin \phi_h}_{LU}$ amplitude of $\pi^0$ from exclusive and non-exclusive two-pion production
exhibits a similar magnitude and similar kinematic dependences as for semi-inclusive $\pi^0$ production. Therefore, a possible contribution of this
exclusive channel to our semi-inclusive result would not alter the presented
asymmetry amplitudes.

\

\

\noindent {\bf Acknowledgement}

\noindent  We thank A. Metz and F.Yuan,  A. Bacchetta, S. Kuhn, A. Kotzinian, A. Prokudin, L. Gamberg and A. Afanasev for useful and stimulating discussions.
We would like to acknowledge the outstanding efforts of the staff of the 
Accelerator and the Physics Divisions at JLab that made this experiment possible.
This work was supported in part by the U.S. Department of Energy
and the National  Science Foundation, 
the Italian Istituto Nazionale di Fisica Nucleare, the 
 French Centre National de la Recherche Scientifique, 
the French Commissariat \`{a} l'Energie Atomique, 
The Southeastern Universities Research Association (SURA) operates the 
Thomas Jefferson National Accelerator Facility for the United States 
Department of Energy under contract DE-AC05-84ER40150. 

\section*{References}
\bibliographystyle{iopart-num}
\bibliography{alu}

\end{document}